\documentclass[12pt,titlepage]{article}
\usepackage{amsmath}
\usepackage{amssymb}

\input{tcilatex}

\begin{document}

\title{Inconsistencies of a purported \\
probability current in the Duffin-Kemmer-Petiau theory}
\author{T.R. Cardoso, L.B. Castro\thanks{%
benito@feg.unesp.br }, A.S. de Castro\thanks{%
castro@pesquisador.cnpq.br.} \\
%EndAName
\\
UNESP - Campus de Guaratinguet\'{a}\\
Departamento de F\'{\i}sica e Qu\'{\i}mica\\
12516-410 Guaratinguet\'{a} SP - Brazil}
\date{}
\maketitle

\begin{abstract}
The Duffin-Kemmer-Petiau (DKP) equation with a square step potential is used
in a simple way with polymorphic purposes. It proves adequate to refuse a
proposed new current that is currently interpreted as a probability
current,to show that the Klein paradox does exist in the DKP theory and to
revise other minor misconceptions diffused in the literature.
\end{abstract}

The first-order Duffin-Kemmer-Petiau (DKP) equation has experimented a
renewal of life due to the discovery of a new conserved four-vector current
density \cite{gho1}-\cite{gho4}, whose positive-definite time component
would be a candidate to a probability density, and as a bonus a hope for
avoiding Klein%
%TCIMACRO{\U{b4}}%
%BeginExpansion
\'{}%
%EndExpansion
s paradox for bosons \cite{gho4}. The DKP equation for a boson minimally
coupled to the electromagnetic field is given by%
\begin{equation}
\left( i\beta ^{\mu }D_{\mu }-m\right) \psi =0  \label{dkp}
\end{equation}%
\noindent where the matrices $\beta ^{\mu }$\ satisfy the algebra%
\begin{equation}
\beta ^{\mu }\beta ^{\nu }\beta ^{\lambda }+\beta ^{\lambda }\beta ^{\nu
}\beta ^{\mu }=g^{\mu \nu }\beta ^{\lambda }+g^{\lambda \nu }\beta ^{\mu }
\label{beta}
\end{equation}%
\noindent the covariant derivative is given by $D_{\mu }=\partial _{\mu
}+ieA_{\mu }$ and the metric tensor is $g^{\mu \nu }=\,$diag$\,(1,-1,-1,-1)$%
. The second-order Klein-Gordon and Proca equations are obtained when one
selects the spin-0 and spin-1 sectors of the DKP theory. A well-known
conserved four-current is given by 
\begin{equation}
J^{\mu }=\bar{\psi}\beta ^{\mu }\psi  \label{corrente}
\end{equation}%
\noindent where the adjoint spinor $\bar{\psi}=\psi ^{\dagger }\eta ^{00}$
with%
\begin{equation}
\eta ^{\mu \nu }=\beta ^{\mu }\beta ^{\nu }+\beta ^{\nu }\beta ^{\mu
}-g^{\mu \nu }  \label{etamunu}
\end{equation}
and $\left( \beta ^{\mu }\right) ^{\dagger }=\eta ^{00}\beta ^{\mu }\eta
^{00}$. The time component of this current is not positive definite but it
may be interpreted as a charge density. An alleged new conserved current,
however, is written as \cite{gho1}, \cite{gho3} 
\begin{equation}
S^{\mu }=\bar{\psi}\eta ^{\mu \nu }\psi \,u_{\nu }  \label{corrente s}
\end{equation}%
\noindent Here $u_{\nu }$ is the unity timelike four-velocity of the
observer ($u^{\nu }u_{\nu }=1$). Since $S^{0}=\psi ^{\dagger }\psi \geqslant
0$ in the lab frame, it might be tempting to interpret this alternative
current as a probability current.

In the present work, the simple problem of scattering in a square step
potential, considered as a time-component of the electromagnetic
four-potential, is used to show not only that this new current leads to
inconsistencies sufficient enough to reject it as a true probability
current, but also to show that Klein%
%TCIMACRO{\U{b4}}%
%BeginExpansion
\'{}%
%EndExpansion
s paradox is absent in Ref. \cite{gho4} just because it was not searched
for, and that it is not necessary to refer to limiting cases of smooth
potentials for finding the appropriate boundary conditions for discontinuous
potentials as done in \cite{che}.

Criticism on $S^{\mu }$ has already a precedent. Struyve et al. \cite{stru}
have shown that $S^{\mu }$ is not conserved when the electromagnetic
interaction is introduced in the DKP equation and so $S^{0}$ can not be
interpreted as a particle probability. They found that%
\begin{equation}
\partial _{\mu }\Theta ^{\mu \nu }=\frac{e}{m}F^{\nu \mu }\,J_{\mu }
\label{tensor}
\end{equation}%
where the energy-momentum tensor $\Theta ^{\mu \nu }=\bar{\psi}\eta ^{\mu
\nu }\psi $ and the electromagnetic field tensor $F_{\mu \nu }=\partial
_{\mu }A_{\nu }-\partial _{\nu }A_{\mu }$. Therefore, $S^{\mu }=\Theta ^{\mu
\nu }u_{\nu }$ is not conserved for an arbitrary observer. Struyve et al. 
\cite{stru} \ argued that the energy-momentum tensor is not conserved
because a charged particle exchanges energy and momentum with the
electromagnetic field. Recently, Datta \cite{dat} has tried to save the
status of the new current by considering%
\begin{equation}
S^{\prime \mu }=S^{\mu }+\frac{e}{m}\int dx^{\mu }F^{\lambda \nu
}\,J_{\lambda }u_{\nu }  \label{newS}
\end{equation}%
with $\partial _{\mu }S^{\prime \mu }=0$. As a matter of fact, Datta tried
to take care of the remark in Ref. \cite{stru} \ about the exchanges of
energy and momentum of the particle with the external field by including the
electromagnetic field as a part of the system.

Let us consider the one-dimensional time component of the static
electromagnetic potential, so that the time-independent DKP equation can be
written as%
\begin{equation}
\{\beta ^{0}[E-eA_{0}(z)]+i\beta ^{3}\frac{d}{dz}-m\}\varphi (z)=0
\label{dkp ind temp}
\end{equation}%
\noindent where the decomposition $\psi (z,t)=\varphi (z)\exp (-iEt)$\ has
been used.

For the case of spin 0, we use the representation for the $\beta ^{\mu }$\
matrices given by \cite{ned1}%
\begin{equation}
\beta ^{0}=%
\begin{pmatrix}
\theta & \overline{0} \\ 
\overline{0}^{T} & \mathbf{0}%
\end{pmatrix}%
\quad \mathrm{and}\quad \beta ^{i}=%
\begin{pmatrix}
\widetilde{0} & \rho _{i} \\ 
-\rho _{i}^{T} & \mathbf{0}%
\end{pmatrix}%
,\quad i=1,2,3  \label{m1}
\end{equation}%
\noindent where%
\begin{eqnarray}
\ \theta &=&%
\begin{pmatrix}
0 & 1 \\ 
1 & 0%
\end{pmatrix}%
\quad \quad \rho _{1}=%
\begin{pmatrix}
-1 & 0 & 0 \\ 
0 & 0 & 0%
\end{pmatrix}
\notag \\
&&  \label{m2} \\
\rho _{2} &=&%
\begin{pmatrix}
0 & -1 & 0 \\ 
0 & 0 & 0%
\end{pmatrix}%
\quad \quad \rho _{3}=%
\begin{pmatrix}
0 & 0 & -1 \\ 
0 & 0 & 0%
\end{pmatrix}
\notag
\end{eqnarray}%
\noindent $\overline{0}$, $\widetilde{0}$ and $\mathbf{0}$ are 2$\times $3, 2%
$\times $2 \ and 3$\times $3 zero matrices, respectively, while the
superscript T designates matrix transposition. The five-component spinor can
be written as $\varphi ^{T}=\left( \varphi _{1},...,\varphi _{5}\right) $ in
such a way that the DKP equation decomposes into%
\begin{equation}
\text{O}_{{\verb|KG|}}\,\varphi _{1}=0,\quad \varphi _{2}=\frac{E-eA_{0}}{m}%
\varphi _{1},\quad \varphi _{3}=\varphi _{4}=0,\quad \varphi _{5}=\frac{i}{m}%
\frac{d\varphi _{1}}{dz}  \label{esp1}
\end{equation}%
\noindent where O$_{{\verb|KG|}}=-d^{2}/dz^{2}+m^{2}-\left( E-eA_{0}\right)
^{2}$. Using now the components of the spinor, the time and space components
of the currents $J^{\mu }$ and $S^{\mu }$ can be written as 
\begin{eqnarray}
J^{0} &=&2\Re\left( \varphi _{1}\varphi _{2}^{\ast }\right) ,\quad
J^{3}=-2\Re(\varphi _{1}\varphi _{5}^{\ast })  \notag \\
&&  \label{corrent} \\
S^{0} &=&\left\vert \varphi _{1}\right\vert ^{2}+\left\vert \varphi
_{2}\right\vert ^{2}+\left\vert \varphi _{5}\right\vert ^{2},\quad
S^{3}=-2\Re\left( \varphi _{2}\varphi _{5}^{\ast }\right)  \notag
\end{eqnarray}%
\noindent Note that there is no reason to require that the spinor and its
derivative are continuous across finite discontinuities of the potential, as
naively advocated in Ref. \cite{che}. A little careful analysis reveals,
though, that proper matching conditions follow from the differential
equations obeyed by the spinor components, as they should be, avoiding in
this manner the hard tasking of recurring to the limit process of smooth
potentials. Only the first component of the spinor satisfies the
second-order Klein-Gordon equation, so that $\varphi _{1}$ and its first
derivative are continuous even the potential suffers finite discontinuities.
In this case of a discontinuous potential, $\varphi _{2}$ is discontinuous
and so are $J^{0}$, $S^{0}$ and $S^{3}$. The discontinuity of $J^{0}$ does
not matter if it is to be interpreted as a charge density. As for $S^{\mu }$%
, it is an obvious nonsense to interpret it as a probability current seeing
that a probability density should always be continuous and that the
probability flux should be uniform in a stationary regime. In this point we
are faced with serious defects of $S^{\mu }$. Nevertheless, despite those
unpleasant properties of $S^{\mu }$ we shall explore the scattering in a
square step potential in order to clarify additional misapprehensions of the
DKP theory.

The one-dimensional square step potential is expressed as 
\begin{equation}
A_{0}\left( z\right) =V_{0}\,\theta \left( z\right)  \label{pot}
\end{equation}%
\noindent where $\theta \left( z\right) $ denotes the Heaviside step
function. For $z<0$ the DKP equation has the solution 
\begin{equation}
\varphi \left( z\right) =\varphi _{+}e^{+ikz}+\varphi _{-}e^{-ikz}
\label{sol z<0}
\end{equation}%
\noindent where%
\begin{equation}
\varphi _{\pm }^{T}=\frac{a_{\pm }}{\sqrt{2}}\left( 1,\frac{E}{m},0,0,\mp 
\frac{k}{m}\right)  \label{sol2}
\end{equation}%
\noindent and $k=\sqrt{E^{2}-m^{2}}$. For $\left\vert E\right\vert >m$, the
solution expressed by (\ref{sol z<0}) and (\ref{sol2}) describes plane waves
propagating on both directions of the $Z$-axis with group velocity $%
v_{g}=dE/dk$ equal to the classical velocity. If we choose particles
inciding on the potential barrier $\left( E>m\right) $, $\varphi _{+}\exp
(+ikz)$ will describe incident particles ($v_{g}=+k/E>0$), whereas $\varphi
_{-}\exp (-ikz)$ will describe reflected particles ($v_{g}=-k/E<0$). The
flux related to the standard current $J^{\mu }$, corresponding to $\varphi $
given by (\ref{sol z<0}), is expressed as 
\begin{equation}
J^{3}=\frac{k}{m}\left( \left\vert a_{+}\right\vert ^{2}-\left\vert
a_{-}\right\vert ^{2}\right)  \label{j1}
\end{equation}%
\noindent Note that the relation $J^{3}=J^{0}v_{g}$ maintains for the
incident and reflected waves, since%
\begin{equation}
J_{\pm }^{0}=\frac{E}{m}\left\vert a_{\pm }\right\vert ^{2}  \label{j2}
\end{equation}%
\noindent On the other hand, for $z>0$ one should have $v_{g}\geq 0$ in such
a way that the solution in this region of space describes an evanescent wave
or a progressive wave running away from the potential interface. The general
solution has the form 
\begin{equation}
\varphi _{\text{t}}\left( z\right) =\left( \varphi _{\text{t}}\right)
_{+}e^{+iqz}+\left( \varphi _{\text{t}}\right) _{-}e^{-iqz}  \label{sol z0}
\end{equation}%
\noindent where%
\begin{equation}
\left( \varphi _{\text{t}}\right) _{\pm }^{T}=\frac{b_{\pm }}{\sqrt{2}}%
\left( 1,\frac{E-eV_{0}}{m},0,0,\mp \frac{q}{m}\right)  \label{sol3}
\end{equation}%
\noindent and $q=\sqrt{\left( E-eV_{0}\right) ^{2}-m^{2}}$. Due to the
twofold possibility of signs for the energy of a stationary state, the
solution involving $b_{-}$ can not be ruled out a priori. As a matter of
fact, this term may describe a progressive wave with negative energy and
phase velocity $v_{ph}=|E|/q>0$. One can readily envisage that three
different classes of solutions can be segregated:

\begin{itemize}
\item \textbf{Class A.} For $eV_{0}<E-m$ one has $q\in 
%TCIMACRO{\U{211d} }%
%BeginExpansion
\mathbb{R}
%EndExpansion
$, and the solution describing a plane wave propagating in the positive
direction of the $Z$-axis with group velocity $v_{g}=q/\left(
E-eV_{0}\right) $ is possible only if $b_{-}=0$. In this case the components
of the standard current are given by
\end{itemize}

\begin{equation}
J^{0}=\frac{E-eV_{0}}{m}\left\vert b_{+}\right\vert ^{2},\quad J^{3}=\frac{q%
}{m}\left\vert b_{+}\right\vert ^{2}  \label{c11}
\end{equation}

\begin{itemize}
\item \textbf{Class B.} For $E-m<eV_{0}<E+m$ one has that $q=\pm i\left\vert
q\right\vert $, and (\ref{sol z0}) with $b_{\mp }=0$ describes an evanescent
wave. The condition $b_{\mp }=0$ is necessary for furnishing a finite
current as $z\rightarrow \infty $. In this case\bigskip 
\begin{equation}
J^{0}=\frac{E-eV_{0}}{m}e^{-2\left\vert q\right\vert z}\left\vert b_{\pm
}\right\vert ^{2},\quad J^{3}=0  \label{c2}
\end{equation}

\item \textbf{Class C.} With $eV_{0}>E+m$ it appears again the possibility
of propagation in the positive direction of the $Z$-axis, now with $b_{+}=0$
and a group velocity given by $v_{g}=q/\left( eV_{0}-E\right) $. The
standard current takes the form%
\begin{equation}
J^{0}=\frac{E-eV_{0}}{m}\left\vert b_{-}\right\vert ^{2},\quad J^{3}=-\frac{q%
}{m}\left\vert b_{-}\right\vert ^{2}  \label{c33}
\end{equation}%
In this last class we meet a bizarre circumstance as long as both $J^{0}$
and $J^{3}$ are negative quantities. The maintenance of the relation $%
J^{3}=J^{0}v_{g}$, though, is a license to interpret the solution $\left(
\varphi _{\text{t}}\right) _{-}\exp (-iqz)$ as describing the propagation,
in the positive direction of the $Z$-axis, of particles with electric
charges of opposite sign to the incident particles. This interpretation is
consistent if the particles moving in this region have energy $-E$ and are
under the influence of a potential $-eV_{0}$. It means that, in fact, the
progressive wave describes the propagation of antiparticles in the positive
direction of the $Z$-axis.
\end{itemize}

\noindent The demand for continuity of $\varphi _{1}$ and $d\varphi _{1}/dz$
at $z=0$ fixes the wave amplitudes in terms of the amplitude of the incident
wave, viz. \bigskip 
\begin{equation}
\frac{a_{-}}{a_{+}}=\left\{ 
\begin{array}{c}
\frac{k-q}{k+q} \\ 
\\ 
\frac{\left( k-i|q|\right) ^{2}}{k^{2}+|q|^{2}} \\ 
\\ 
\frac{k+q}{k-q}%
\end{array}%
\begin{array}{c}
\text{\textrm{for the class}}{\text{ }\mathbf{A}} \\ 
\\ 
\text{\textrm{for the class}}{\text{ }\mathbf{B}} \\ 
\\ 
\text{\textrm{for the class}}{\text{ }\mathbf{C}}%
\end{array}%
\right.  \label{12}
\end{equation}%
\begin{equation}
\frac{b_{+}}{a_{+}}=\left\{ 
\begin{array}{c}
\frac{2k}{k+q} \\ 
\\ 
\frac{2k\left( k-i|q|\right) }{k^{2}+|q|^{2}} \\ 
\\ 
0%
\end{array}%
\begin{array}{c}
\text{\textrm{for the class}}{\text{ }\mathbf{A}} \\ 
\\ 
\text{\textrm{for the class}}{\text{ }\mathbf{B}} \\ 
\\ 
\text{\textrm{for the class}}{\text{ }\mathbf{C}}%
\end{array}%
\right.  \label{13}
\end{equation}%
\begin{equation}
\frac{b_{-}}{a_{+}}=\left\{ 
\begin{array}{c}
0 \\ 
\\ 
0 \\ 
\\ 
\frac{2k}{k-q}%
\end{array}%
\begin{array}{c}
\text{\textrm{for the class}}{\text{ }\mathbf{A}} \\ 
\\ 
\text{\textrm{for the class}}{\text{ }\mathbf{B}} \\ 
\\ 
\text{\textrm{for the class}}{\text{ }\mathbf{C}}%
\end{array}%
\right.  \label{14}
\end{equation}

Now we focus attention on the calculation of the reflection ($R$) and
transmission ($T$) coefficients. The reflection (transmission) coefficient
is defined as the ratio of the reflected (transmitted) flux to the incident
flux. Since $\partial J^{0}/\partial t=0$ for stationary states, one has
that $J^{3}$ is independent of $z$. This fact implies that \bigskip 
\begin{equation}
R=\left\{ 
\begin{array}{c}
\left( \frac{k-q}{k+q}\right) ^{2} \\ 
\\ 
1 \\ 
\\ 
\left( \frac{k+q}{k-q}\right) ^{2}%
\end{array}%
\begin{array}{c}
\text{\textrm{for the class}}{\text{ }\mathbf{A}} \\ 
\\ 
\text{\textrm{for the class}}{\text{ }\mathbf{B}} \\ 
\\ 
\text{\textrm{for the class}}{\text{ }\mathbf{C}}%
\end{array}%
\right.  \label{15}
\end{equation}
\begin{equation}
T=\left\{ 
\begin{array}{c}
\frac{4kq}{(k+q)^{2}} \\ 
\\ 
0 \\ 
\\ 
-\frac{4kq}{(k-q)^{2}}%
\end{array}%
\begin{array}{c}
\text{\textrm{for the class}}{\text{ }\mathbf{A}} \\ 
\\ 
\text{\textrm{for the class}}{\text{ }\mathbf{B}} \\ 
\\ 
\text{\textrm{for the class}}{\text{ }\mathbf{C}}%
\end{array}%
\right.  \label{16}
\end{equation}

\noindent For all the classes one has \ $R+T=1$ as should be expected for a
conserved quantity. The class C presents $R>1$, the alluded Klein%
%TCIMACRO{\U{b4}}%
%BeginExpansion
\'{}%
%EndExpansion
s paradox, implying that more particles are reflected from the potential
barrier than those incoming. Contrary to the assertion of Ghose et al. \cite%
{gho4}, Klein%
%TCIMACRO{\U{b4}}%
%BeginExpansion
\'{}%
%EndExpansion
s paradox there exists for bosons in the DKP theory. It must be so because,
as seen before, the potential stimulates the production of antiparticles at $%
z=0$. Due to the charge conservation there is, in fact, the creation of
particle-antiparticle pairs. Since the potential in $z>0$ is repulsive for
particles they are necessarily reflected. From the previous discussion
related to the classes B and C, one can realize that the threshold energy
for the pair production is given by $eV_{0}=2m$. The propagation of
antiparticles inside the potential barrier can be interpreted as due to the
fact that each antiparticle is under the influence of an effective potential
given by $-eV_{0}$. In this way, each antiparticle has an available energy
(rest energy plus kinetic energy) given by $eV_{0}-E$, accordingly one
concludes about the threshold energy. One can also say that there is an
ascending step for particles and a descending step for antiparticles.

Note that the currents $J^{\mu }$ and $S^{\mu }$ are simply related by%
\begin{equation}
S^{\mu }=\frac{E-eA_{0}}{m}\,J^{\mu }  \label{rel}
\end{equation}%
\noindent in all the classes of solutions. In this manner, the conservation
law $\partial _{\mu }J^{\mu }=0$ is not compatible with $\partial _{\mu
}S^{\mu }=0$, at least for the case under investigation where $\partial
S^{0}/\partial t=0$ but $S^{3}$ is not uniform. In order to understand the
behaviour of $S^{\mu }$ let us recall that the DKP equation can be recast
into the Hamiltonian form \cite{now}, \cite{pim1} 
\begin{equation}
i\frac{\partial \psi }{\partial t}=H\psi  \label{eqham}
\end{equation}%
\noindent where 
\begin{equation}
H=i\left[ \beta ^{i},\beta ^{0}\right] D_{i}+eA_{0}+m\beta ^{0}+i\frac{e}{2m}%
F_{\mu \nu }\left( \beta ^{\mu }\beta ^{0}\beta ^{\nu }+\beta ^{\mu }g^{0\nu
}\right)  \label{ham}
\end{equation}%
\noindent At this point is worthwhile to mention that $H$ is not Hermitian,
in opposition what was adverted in \cite{now}, since 
\begin{equation}
\left( iF_{0i}\beta ^{i}\left( \beta ^{0}\right) ^{2}\right) ^{\dagger
}=-\left( iF_{0i}\beta ^{i}\left( \beta ^{0}\right) ^{2}\right)
+iF_{0i}\beta ^{i}  \label{her}
\end{equation}%
There results that 
\begin{equation}
\partial _{\mu }S^{\mu }=i\psi ^{\dagger }\left( H-H^{\dagger }\right) \psi
=-\frac{e}{m}F_{0i}\,\psi ^{\dagger }\left[ \beta ^{i},\left( \beta
^{0}\right) ^{2}\right] \psi =\frac{e}{m}F_{0i}\,J^{i}  \label{source1}
\end{equation}%
\noindent This result clearly shows that the electromagnetic coupling
induces a source term in the current $S^{\mu }$, as has already been shown
in Ref. \cite{stru}. It is curious that the source term is due to the
non-Hermitian piece of the anomalous term in (\ref{ham}). Now, coming back
to the square step potential (\ref{pot}), one can write 
\begin{equation}
\partial _{\mu }S^{\mu }=-\frac{eV_{0}}{m}\,\delta \left( z\right)
J^{3}\left( z\right)  \label{source2}
\end{equation}%
\noindent in such a way that the jumping of $S^{3}$ at $z=0$ reads 
\begin{equation}
S^{3}(0_{+})-S^{3}(0_{-})=-\frac{eV_{0}}{m}J^{3}(0)  \label{source3}
\end{equation}%
\noindent a result in perfect agreement with (\ref{rel}).

As for the current proposed by Datta \cite{dat}, one can see that there is a
spurious factor of 4 in the second term of $S^{\prime \mu }$ in (\ref{newS})
due to the process of summation of four identical terms (four uses of the
fundamental theorem of calculus involving $F^{\mu \nu }\,J_{\mu }u_{\nu }$)
for considering the source term in (\ref{tensor}) as an additional current
term, not this merely but also $\partial _{\mu }S^{\prime \mu }\neq 0$. The
new conserved current should be written as 
\begin{equation}
\tilde{S}^{\mu }=S^{\mu }+\frac{1}{4}\frac{e}{m}\int dx^{\mu }F^{\lambda \nu
}\,J_{\lambda }u_{\nu }
\end{equation}%
so that $\tilde{S}^{\mu }$ without any question satisfies $\partial _{\mu }%
\tilde{S}^{\mu }=0$. It is worthwhile to note, though, that all components
of the conserved current are nonvanishing for an arbitrary direction of
motion. Furthermore, $\tilde{S}^{0}$ carries a temporal dependence even for
the case of a time-independent DKP equation and it is infinite at the points
of space where the potential suffers finite discontinuities. These are a few
undesirable features which defy the candidature of $\tilde{S}^{\mu }$ as a
current probability.

For short, the DKP equation with a square step potential is a test ground to
refuse $S^{\mu }$ (and $\tilde{S}^{\mu }$) as a probability current as well
as to show that Klein%
%TCIMACRO{\U{b4}}%
%BeginExpansion
\'{}%
%EndExpansion
s paradox is alive and well in the DKP theory (we have talking about the
spin-0 sector of the DKP theory but the state of affairs is not different
for the spin-1 sector as one can see in Appendix A).

\bigskip

\bigskip

\noindent\textbf{Appendix A}

For the case of spin 1, the $\beta ^{\mu }$\ matrices are \cite{ned2}%
\begin{equation}
\beta ^{0}=%
\begin{pmatrix}
0 & \overline{0} & \overline{0} & \overline{0} \\ 
\overline{0}^{T} & \mathbf{0} & \mathbf{I} & \mathbf{0} \\ 
\overline{0}^{T} & \mathbf{I} & \mathbf{0} & \mathbf{0} \\ 
\overline{0}^{T} & \mathbf{0} & \mathbf{0} & \mathbf{0}%
\end{pmatrix}%
\quad \quad \beta ^{i}=%
\begin{pmatrix}
0 & \overline{0} & e_{i} & \overline{0} \\ 
\overline{0}^{T} & \mathbf{0} & \mathbf{0} & -is_{i} \\ 
-e_{i}^{T} & \mathbf{0} & \mathbf{0} & \mathbf{0} \\ 
\overline{0}^{T} & -is_{i} & \mathbf{0} & \mathbf{0}%
\end{pmatrix}%
\end{equation}%
\noindent where $s_{i}$ are the 3$\times $3 spin-1 matrices $\left(
s_{i}\right) _{jk}=-i\varepsilon _{ijk}$, $e_{i}$ are the 1$\times $3
matrices $\left( e_{i}\right) _{1j}=\delta _{ij}$ and $\overline{0}=%
\begin{pmatrix}
0 & 0 & 0%
\end{pmatrix}%
$, while\textbf{\ }$\mathbf{I}$ and $\mathbf{0}$\textbf{\ }designate the 3$%
\times $3 unit and zero matrices, respectively. With the spinor written as $%
\varphi ^{T}=\left( \varphi _{1},...,\varphi _{10}\right) $, and defining $%
\Psi ^{T}=\left( \varphi _{2},\varphi _{3},\varphi _{7}\right) $, $\Phi
^{T}=\left( \varphi _{5},\varphi _{6},\varphi _{4}\right) $and $\Theta
^{T}=\left( \varphi _{9},-\varphi _{8},\varphi _{1}\right) $ as done in Ref. 
\cite{che}, the DKP equation (\ref{dkp ind temp}) can be expressed in terms
of the following equations%
\begin{equation}
\text{O}_{{\verb|KG|}}\Psi =0,\quad \Phi =\frac{E-eV_{0}}{m}\,\Psi ,\quad
\Theta =\frac{i}{m}\frac{d\Psi }{dz},\quad \varphi _{10}=0  \label{spin1a}
\end{equation}%
\noindent Now the components of the four-currents are given by%
\begin{eqnarray}
J^{0} &=&2\Re\left( \varphi _{2}\varphi _{5}^{\ast }+\varphi _{3}\varphi
_{6}^{\ast }+\varphi _{4}\varphi _{7}^{\ast }\right) ,\quad
J^{3}=-2\Re\left( \varphi _{1}\varphi _{7}^{\ast }+\varphi _{2}\varphi
_{9}^{\ast }-\varphi _{3}\varphi _{8}^{\ast }\right)  \notag \\
&&  \label{spin1b} \\
S^{0} &=&\sum_{i=1}^{9}|\varphi _{i}|^{2},\quad S^{3}=-2\Re\left( \varphi
_{1}\varphi _{4}^{\ast }+\varphi _{5}\varphi _{9}^{\ast }-\varphi
_{6}\varphi _{8}^{\ast }\right)  \notag
\end{eqnarray}%
\noindent A discontinuous potential makes $\varphi _{4}$, $\varphi _{5}$ and 
$\varphi _{6}$ discontinuous, and as an immediate consequence $J^{0}$, $%
S^{0} $ and $S^{3}$ are also discontinuous. The plane wave solutions for the
potential given by (\ref{pot}) in the region $z<0$ can be written as%
\begin{equation}
\varphi \left( z\right) =\varphi _{+}e^{+ikz}+\varphi _{-}e^{-ikz}
\label{spin1c}
\end{equation}%
\noindent where%
\begin{equation}
\varphi _{\pm }^{T}=\left( \mp \frac{k}{m}\gamma _{\pm },\alpha _{\pm
},\beta _{\pm },\frac{E}{m}\gamma _{\pm },\frac{E}{m}\alpha _{\pm },\frac{E}{%
m}\beta _{\pm },\gamma _{\pm },\pm \frac{k}{m}\beta _{\pm },\mp \frac{k}{m}%
\alpha _{\pm },0\right)  \label{spin1d}
\end{equation}%
\noindent and $\alpha _{\pm }$, $\beta _{\pm },$and $\gamma _{\pm }$ are
arbitrary amplitudes. Defining%
\begin{equation}
a_{\pm }=\sqrt{2\left( |\alpha _{\pm }|^{2}+|\beta _{\pm }|^{2}+|\gamma
_{\pm }|^{2}\right) }
\end{equation}%
\noindent it follows that the components of the current can be written in
the same form as (\ref{j1}) and (\ref{j2}). A similar procedure for the
region $z>0$ allows one to obtain the results (\ref{12})-(\ref{16}). Even
though $\varphi _{2}$, $\varphi _{3}$ and $\varphi _{7}$ obey the
Klein-Gordon equation there is no reason why they have the same amplitudes,
as assumed in Ref. \cite{che}. As a matter of fact, a nontrivial spinor with
only three nonvanishing components would be possible.

\bigskip

\bigskip

\bigskip

\noindent \textbf{Acknowledgments}

This work was supported in part by means of funds provided by CAPES, CNPq
and FAPESP. The authors would like to thank Professor P. Holland for drawing
attention to Ref. \cite{stru} after submission of our Letter to Physics
Letters A and to an anonymous referee for Ref. \cite{dat}.

\medskip

\end{document}